\documentclass[useAMS,usenatbib]{mn2e}
\usepackage{textcomp,graphicx,amsmath,pdflscape,amssymb,subfigure}
\usepackage{epsfig}
\usepackage{amsmath}
\usepackage{times} 
\usepackage{epsf}
\usepackage{lscape}
\usepackage{rotating}
\usepackage{multirow}

\newcommand\apj{{ApJ}}%
\newcommand\apjl{{ApJ}}%
\newcommand\apss{{Ap\&SS}}%
\newcommand\aap{{A\&A}}%
\newcommand\aaps{{A\&AS}}%
\newcommand\mnras{{MNRAS}}%
\begin{document}

\title[Directed Follow-Up approach for \textit{Gaia} photometry]
{Directed follow-up strategy of low-cadence photometric surveys in Search of Transiting Exoplanets - 
II. application to \textit{Gaia}}

\author[Y. Dzigan and S. Zucker]
  {Yifat Dzigan$^1$\thanks{E-mail: yifatdzigan@gmail.com}
  and Shay Zucker$^1$\thanks{E-mail: shayz@post.tau.ac.il}\\
$^1$Department of Geophysics, Atmospheric and Planetary Sciences, Tel Aviv University, Tel Aviv 69978, Israel\\
 }
\date{}
\pagerange{\pageref{firstpage}--\pageref{lastpage}} \pubyear{2012}

\maketitle
\label{firstpage}

\begin{abstract}

In a previous paper we presented the Directed Follow-Up (DFU) approach, which we
suggested can be used to efficiently augment low-cadence photometric surveys in
a way that will optimize the chances to detect transiting exoplanets. In this paper we 
present preliminary tests of applying the DFU approach to the future
ESA space mission \textit{Gaia}. We demonstrate the strategy application to
\textit{Gaia} photometry through a few simulated cases of known transiting
planets, using \textit{Gaia} expected performance and current design. We show
that despite the low cadence observations DFU, when tailored for \textit{Gaia}'s
scanning law, can facilitate detection of transiting planets with ground-based
observations, even during the lifetime of the mission. We conclude that
\textit{Gaia} photometry, although not optimized for transit detection, should
not be ignored in the search of transiting planets. With a suitable ground-based
follow-up network it can make an important contribution to this search.

\end{abstract}


\section{Introduction}\label{intro}

\textit{Gaia} is a European Space Agency (ESA) mission, scheduled to
be launched at 2013. It will provide astrometric, spectroscopic and
photometric measurements of the sky between the 6th and 20th magnitude
\citep{2009sf2a.conf...45E}. As the successor of \textit{Hipparcos},
\textit{Gaia} is supposed to improve on the accuracy of
\textit{Hipparcos}. Specifically in photometry, it is expected to
reach a milli-magnitude ($\mathrm{mmag}$) photometric precision down
to the 16th $G$-magnitude \citep{2012Ap&SS.tmp...68D}. Nominally, this
may enable the photometric detection of planetary transits
\citep{2012ApJ...753L...1D}.

According to our recent estimates \citep{2012ApJ...753L...1D}, the
potential yield of transiting planets from \textit{Gaia} photometry
can reach a few thousands transiting exoplanets, depending on the
number of planetary transits that the telescope should sample to
secure detection. Due to the low cadence of the observations,
\textit{Gaia} will typically sample very few transits, which implies
the need for a detection algorithm that will be tailored for
\textit{Gaia} special features, and will require a minimal number of
in-transit observations.

In a previous paper \citep{2011MNRAS.415.2513D} we first presented a
strategy, which we named Directed Follow-Up (DFU), that is suitable
for the detection of transiting exoplanets in low-cadence surveys.
DFU uses specially tailored ground-based follow-up observations to
augment the low-cadence data.  In this paper we use \textit{Gaia}
scanning law to simulate light curves of transiting planets, and
examine the capability of our strategy to detect transits in
\textit{Gaia}'s photometry and to predict optimal timing of follow-up
observation.

\subsection{The \textit{Gaia} mission}\label{window_func}

The primary goal of the \textit{Gaia} space mission is to explore the
formation, dynamical, chemical and star-formation evolution of the
Milky Way galaxy. It will achieve those goals using high precision
astrometry, that will be backed with photometry and
spectroscopy. During it's five-year run, \textit{Gaia} will measure
parallaxes, luminosities, and proper motions for $\sim 10^9$ stars in
our Galaxy and beyond, as well as quasars, Solar System objects and
other galaxies \citep{2010IAUS..261..296L}.

\textit{Gaia} will operate in a Lissajous-type orbit, around
the Sun--Earth L2 point, so it will corotate with Earth in its orbit
around the Sun, at a distance of about $1.5$ million kilometers from
Earth, in the anti-Sun direction. The mission will have a dual
telescope, with a common structure and common focal plane. The
spacecraft will rotate around an axis that will be perpendicular to
the two fields of view, with a constant rate of $1\degr$ per minute,
in order to repeat the observations in the two fields of view.  Due to
the basic angle of $106.5\degr$ that will separate the astrometric
fields of view on the sky, objects will transit the two fields with a
delay of $106.5$ minutes \citep{2012Ap&SS.tmp...68D}. The spacecraft's
spin motion of six-hour period, and a $63$-day-period precession will
cause \textit{Gaia} scanning law to be peculiar and irregular. This
special scanning law will result in an average of $70$ measurements
per object.

\textit{Gaia} will provide photometry in several passbands. The $BP$
($330$--$680$~nm) and $RP$ ($640$--$1050$~nm) bands correspond to the
blue and red \textit{Gaia} photometers. Those photometers will provide
low-resolution spectrophotometric measurements. A third passband will
be a wide and 'white' passband dubbed $G$, that will be centred on
$\lambda_0 = 673\,\mathrm{nm}$, with a width of $\Delta\lambda =
440\,\mathrm{nm}$.  One can expect a $\mathrm{mmag}$ precision in the
$G$ band, down to the 14th--16th $G$ magnitude, and
$10\,\mathrm{mmag}$ for the faintest objects
\citep{2010A&A...523A..48J}. The exact limiting magnitude for a
$1\,\mathrm{mmag}$ precision depends on the final observing strategy
and on instrumental factors, which are not yet fully determined
\citep{2012Ap&SS.tmp...68D}.

Estimates of the expected yield of transiting planets from
\textit{Gaia} photometry range from hundreds to thousands
\citep[e.g.,][]{2002Ap&SS.280..139H, 2002EAS.....2..215R}. In our
previous paper \citep{2012ApJ...753L...1D} we estimated the expected
yield, based on a statistical methodology, presented by
\cite{2008ApJ...686.1302B}. We used assumptions regarding the
galactic structure, effects of stellar variability, and transiting
planet frequencies based on complete transit surveys (namely OGLE). As
could be expected, our results suggested that \textit{Gaia}'s
transiting planet yield will depend strongly on the transit sampling
(number of distinct transits that the telescope will sample per
system). This intuitive result can be quantified by the observational
window function \citep{2009ApJ...702..779V}, which is a function of
the planet orbital period.

\begin{figure}
\includegraphics[width=0.5\textwidth]{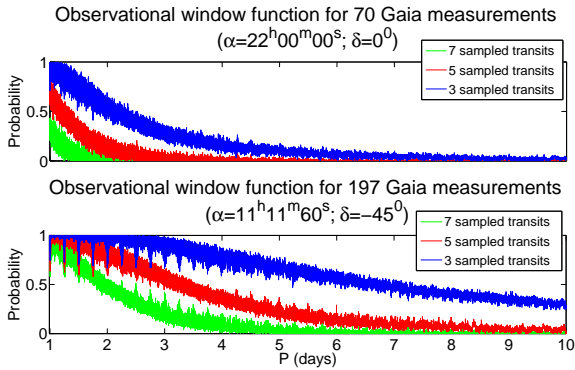}
\caption{Observational window functions for two areas of
\textit{Gaia}'s sky.
\textit{Top}: A sky direction that \textit{Gaia} is expected to visit
$70$ times, which is the average number of \textit{Gaia} measurements
over the five year time span of the mission. A similar window function was presented in 
\citet{2012ApJ...753L...1D}.  \textit{Bottom}: A sky
direction that \textit{Gaia} is expected to sample $197$ times. Both
window functions were calculated for a transit duration of
$2\,\mathrm{hr}$, for a minimum of three, five and seven in-transit
observations.}
\label{fig.window_funcion1}
\end{figure}

Assuming the existence of a transiting planet, the probability to
sample a minimum of three, five and seven transits for a typical
\textit{Gaia} star, with $70$ measurements, is shown in the upper panel
of Fig.~\ref{fig.window_funcion1}, together with another case, less
probable, with $197$ measurements, in the bottom panel. 

The sharp
minima and maxima in the window function for the extreme case are due
to the six-hour-period spin of the telescope
\citep{2005MNRAS.361.1136E}.  For the case of a typical \textit{Gaia}
light curve with $\sim70$ measurements, if we assume the existence of
a transiting Hot Jupiter (HJ), the probability to sample at least
seven separate transits is practically negligible. The probability to
sample a minimum of five transits is $\sim5$ per cent, but the
probability to sample a minimum of three transits increases to
$\sim30$ per cent. Thus, it may prove very beneficial to somehow relax
the requirement for a minimum number of sampled transits.  The DFU
strategy aims to achieve exactly that.

\subsection{Directed Follow-Up}\label{dfu}

The DFU strategy is based, in principle, on Bayesian inference, (e.g.,
implemented by a Markov-Chain Monte-Carlo -- MCMC -- procedure), that
we use to estimate the posterior probability density functions of the
transit parameters. We assume that a simple box-shape transit
light-curve model describes the data
\citep[.e.g,][]{2002A&A...391..369K}. This model involves five
parameters: the period -- $P$, phase -- $T_c$, and width of the
transit -- $w$, and the flux levels in-transit and ex-transit. The
MCMC procedure we chose to use was the Metropolis-Hastings (MH)
algorithm
\citep{2011MNRAS.415.2513D}.

We first apply the MH algorithm to the measurements of a target
star. The results (after excluding the appropriate `burning time') are
Markov chains of the successful iterations, for each of the model
parameters. From each chain we extract the stationary distribution of
the parameter, which we use as the estimated Bayesian posterior
distribution \citep{2005blda.book.....G}.  If the low-cadence data
happen to sample enough separate planetary transits, with sufficient
precision, we expect the distributions to concentrate around the
actual values of the parameters. Otherwise, If the data samples a
small number of transits, the distributions will probably spread over
different solutions, besides the unknown actual one.

The second step of the strategy is assigning to each point in time the
probability that a transit will occur at that time. We calculate the
transit probability using the posterior distributions found by the MH
algorithm. For each point in time, we count the number of successful
iterations whose values of $P$, $T_c$ and $w$ predict that a transit
will occur at the examined time.  This results in a function we termed
the Instantaneous Transit Probability -- ITP.

An important part of the DFU strategy is the choice of stars to
follow. We propose to choose stars for follow-up observations
according to three criteria:

The first criterion follows from our observation that the most robust
indication that a periodic transit-like signal exists in the data is
the maximum value of the ITP. This should be intuitive if one
remembers that by definition the ITP peak values actually represent
the probability to sample a transit in follow-up observations.

The second criterion we propose is the skewness of the ITP, defined as
\begin{equation}
 S=\frac{\langle (x-\langle \hat{x}\rangle)^3\rangle}{\sigma^3} \ ,
\end{equation}
where $x$ represents the ITP function values, and $\langle \cdot
\rangle$ denotes the averaging operation.  In a sense, the skewness
measures the amount of outliers in the ITP function, where `outliers'
refers here to ITP prominently high values. The relative absence of
outliers, i.e., a 'flat' ITP, will yield a skewness that is close to
zero ($S=0$ for a Gaussian distribution), which means that no time is
preferred for follow-up observations.

The third criterion we propose is the Wald statistic of the transit
depth posterior distribution. This simply quantifies the significance
of the transit depth by measuring it in terms of its own standard
deviation.  Our experience shows that in case the algorithm explores
the transit depth parameter space without converging, it might yield a
high Wald statistic value, which we might erroneously interpret as
evidence of a transit signal. Therefore, we propose to use this
criterion only for prioritizing stars that have passed the first two
criteria.

The third and final step of the strategy is to actually perform
follow-up observations, at the times preferred by the ITP, thus
optimizing the chances to detect the transit in a few observations as
possible.  We then propose to combine both the `old' data from the
survey and the new observations, to recalculate the new posterior
distributions that reflect our new state of knowledge, and to propose
new times for the next follow-up observations.

In the most favorable case, where a follow-up observation happens to
actually take place during transit, it will usually eliminate most
spurious peaks in the period posterior distribution (PPD), except for
the actual orbital period of the planet. Then a careful high-cadence
photometric or spectroscopic follow-up of the candidates can confirm
their planetary nature.  In case we do \textit{not} observe the
transit in the follow-up observations, the new data will also
eliminate some periods that will not fit our new state of knowledge,
once again, resulting in new posterior distributions of the model
parameters.  The whole procedure should be repeated until the
detection of a transiting planet, or, alternatively the exclusion of
its existence (to a specified degree of certainty).

\section{DFU for \textit{Gaia}}\label{Gaia_application}

We propose to apply the DFU strategy to the \textit{Gaia} data in the following way:

\noindent
1. Select the initial targets.

\noindent
2. Run the MH algorithm for the targets that were selected in step 1.

\noindent
3. Compute the Instantaneous Transit Probability function.

\noindent
4. Prioritize stars for directed follow-up observations according to the ITP peak value and skewness.

\noindent
5. Perform follow-up observations of the selected stars, and then combine
the new data with \textit{Gaia}'s data to repeat the process, starting
from step 2.

In the first stage, when selecting the targets, one can obviously use
various astrophysical criteria, e.g., the metallicity, which might
affect the apriori probability of the star to host planets. This step
can use the whole \textit{Gaia} dataset with its various components of
astrometric and spectroscopic information . Using the astrophysical
and astronomical considerations, one will probably also eliminate
beforehand eclipsing binaries, known variable stars, and
non-main-sequence stars.  We also propose to run a test for the
possible presence of a transit signal in the data, by examining the
statistical distribution of the light-curve values. Ideally, those
values will have a bimodal asymmetric distribution, whose two peaks
correspond to the in-transit and ex-transit levels, with corresponding
relative probabilities. The plausibility of this distribution may
serve as a preliminary screening tool, before applying the
computationally demanding MH algorithm.

Once one has selected the targets and applied the MH algorithm to the
light curves of the chosen targes, DFU can culminate in detection
through one of three possible scenarios. The \textit{first scenario}
is a detection of the transit using solely \textit{Gaia}
photometry. We define detection as a case in which the MH algorithm
results in a PPD that is centred around a single solution, (and maybe
its harmonics and sub-harmonics).  This very narrow distribution will
ensure that the resulting ITP peaks will coincide with actual
transits, with high detection probability, and that we will be able to
use those preferred times to conduct high cadence follow-up
observations, to confirm the transit nature of the periodic signal and
refine its parameters.

Due to \textit{Gaia}'s low cadence, and relatively small number of
observations per star, in most realistic cases we expect to obtain
posterior distributions that will exhibit several possible periods
\citep{2011MNRAS.415.2513D}.  This will result in the other two more
probable scenarios:

If the PPD exhibits a small number of peaks (that are not
harmonics of each other), we will face the \textit{second scenario},
in which a single follow-up observation will be needed in order to
finally detect the transit. In the \textit{third scenario}, the PPD
will be distributed over many peaks, which means it will require more
than a single follow-up observation in order to finally detect the
transit.

\section{simulations}

We simulated light curves of planetary transits according to
\textit{Gaia}'s scanning law and expected photometric precision, as of August 2011. 
We used the same scanning law setup as in \cite{2012ApJ...753L...1D} to assign 
each planet with the timing sequence of \textit{Gaia}
observations, according to its location on the sky. 
Note that the details of the scanning law simulations depends on some arbitrary factors, depending on the exact timing of 
the mission, which is obviously not determined yet. The number of \textit{Gaia} FOV crossings of 
a given object as a function of location in the sky, is 
influenced by the phasing of the scanning law we choose, thus our implementation of the scanning 
law is only representative, and specific attempts of transit 
detection and follow-up should be made using the actual
timing of \textit{Gaia} measurements. 

Following \cite{2012ApJ...753L...1D} we 
assumed one in-transit measurement per transit, for all the simulations we present 
in this paper. It is possible that a transit duration of a few hours will result in more than a single 
FOV crossing per transit, however, several in-transit measurements of the same transit 
do not have significant contribution for constraining the PPD. They can, however, be useful for 
eliminating calibration errors.

We simulated the errors as purely Gaussian. In order to test our
strategy, we examined simulations of known transiting planets with
orbital periods ranging from $\sim1$ to $\sim9$ days. 

The simulations were assigned with different numbers of in-transit observations, by
varying the transit phase, and with a range of transit depths
($d=0.01-0.001\,\mathrm{mag}$) to simulate different planetary sizes. 
Thus, we used the known period and
duration ($w$) of each planet, while varying the phase and depth, according
to the case we aimed to test. Table \ref{table.planet_list} presents
the orbital elements of the known exoplanets we used in the
simulations. 

\begin{figure}
\includegraphics[width=0.5\textwidth]{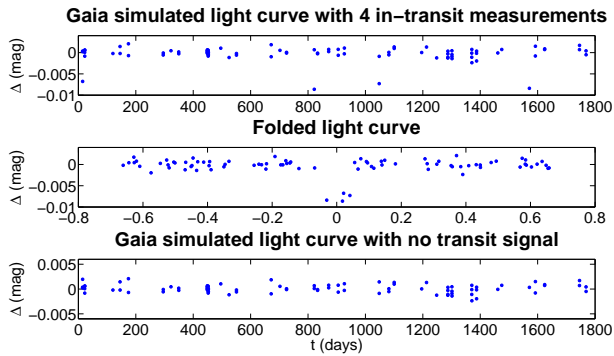}
\caption{\textit{Top}: Simulated \textit{Gaia} light curve of WASP-4b, that corresponds to a 
transit depth of $8\,\mathrm{mmag}$, and four in-transit measurements. \textit{Middle}: 
The simulated light curve of WASP-4b, folded according to the 
transit phase. \textit{Bottom}: Simuated light curve of the same star, 
but with no transit signal.}
\label{fig.light_curve}
\end{figure}

In the upper panel of Fig.~\ref{fig.light_curve} we present an example of a simulated \textit{Gaia} 
light curve of a transit with depth of $8\,\mathrm{mmag}$, to illustrate the challenge to detect 
planetary transit in low-cadence photometric surveys. In the middle panel of Fig.~\ref{fig.light_curve} we 
show the same light curve, folded according to the transit phase, and in the 
bottom panel a light curve of the same star, with no transit signal. 

In some cases, those that demonstrate the second and third detection
scenario, we also simulated the directed follow-up observations,
obtained on times that were indicated optimal for follow-up based on
the ITP values. Each simulated follow-up observing sequence comprised
four single exposures, two that were scheduled to occur within the ITP
peak, and two before and after the peak. This was meant in order to
sample times in and out of the suspected transit. We assumed a
photometric error of $1$\,mmag.

Besides testing the detection using the full simulation of
\textit{Gaia}'s scanning law, we also addressed two additional issues.
The first is detection based on only part of \textit{Gaia}'s full
mission lifetime. \cite{2012IAUS..285..425W} described the
\textit{Gaia} Science Alerts team, that is assigned to handle mainly
transient phenomena in the \textit{Gaia} data stream.  Supernovae,
microlensing events, and m-dwarf flares are just a few examples of the
possible triggers that the team will study.  Our DFU strategy may
benefit from a similar follow-up network that will be able to
follow-up on prominent ITP peaks during the mission lifetime.  To test
the value of such an effort we tested our strategy when only half of
\textit{Gaia} time span was used.

The second issue is the false-alarm rate.  As we have shown in our
previous paper \citep{2012ApJ...753L...1D}, the false-alarm rate due
to \textit{Gaia}'s white Gaussian noise is negligible. However, 
\cite{2012A&A...539A.137M} estimated that more than $60$~per-cent of the stars
exhibit microvariability that is larger than that of the Sun.  Due to
\textit{Gaia}'s low cadence, the effect of the stellar red noise
\citep{2012A&A...539A.137M} is simply an increase in the white noise
level by $1-2\,\mathrm{mmag}$ . We therefore simulated thousands of
\textit{Gaia} light curves with increased Gaussian errors, and
examined the false alarm rate due to stellar variability.

\section{Results}\label{results}

Table \ref{table.table_itp} presents a summary of the test cases we
simulated and the simulation results. The first column is the name of
the planet on which we based the simulation (sky location, period and
transit duration). In some cases, the simulation included also the
directed follow-up observing sequences, which is indicated in the
relevant lines.

The second column is the total number of samples $N_\mathrm{tot}$,
which is determined by the planet coordinates through \textit{Gaia}'s
scanning law. In cases where we tested detection using half of the
mission lifetime, the number is especially low.  Next is the number of
in-transit measurements, $N_\mathrm{tr}$, followed by the transit depth
$d$. The results of the simulation are summarized by the following
columns which are the posterior mean transit depth $\langle d
\rangle$, the Wald statistic $W$, the ITP maximum value and the ITP
skewness $S$.

One immediate result that the table demonstrates is that a transit
depth of $1$\,mmag is simply undetectable. All the detection criteria,
the Wald statistic, the maximum ITP, and the ITP skewness, are
practically similar to what we see when we simulate pure Gaussian
noise. In the deeper transits, $S$ is always larger than $1$, which we
chose as the threshold value for our second criterion.

Below we demonstrate the different detection scenarios we have
mentioned.  For each example
(Fig.~\ref{fig.corot1_itp}--\ref{fig.wasp4_fu}) we present the PPD and
the ITP for various transit depths. We also show some simulations of
directed follow up observations.

\subsection{First detection scenario}\label{detection}

\begin{figure}
\includegraphics[width=0.5\textwidth]{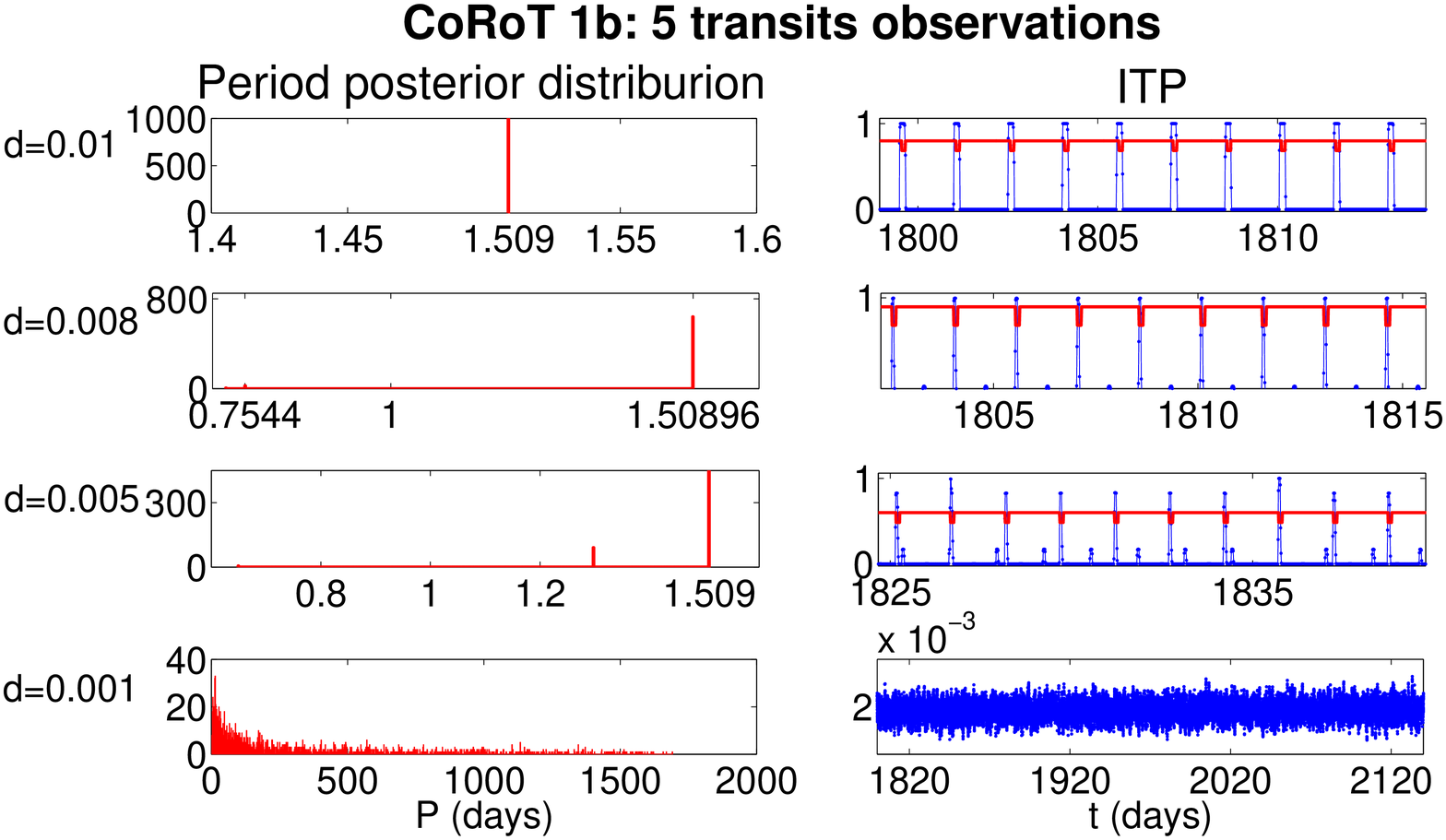}
\caption{PPDs (left panels), and ITP functions (right panels) for the
simulation of CoRoT-1b with five transit observations, for a range of
transit depths.}
\label{fig.corot1_itp}
\end{figure}

We used the planet CoRoT-1b to base on it the first-scenario
simulations, where detection is possible based on \textit{Gaia} data
alone. Fig.~\ref{fig.corot1_itp} shows the PPD and the ITP we
obtained. The choice of the simulated transit phase in such a way that
five transits are sampled allows the detection. We see that for
transits deeper than $d=0.005\,\mathrm{mag}$ (The upper two panels),
the PPDs are centred around a single distinct period, consistent with
the known simulated period. As a result, the ITP prominent peaks
coincide with the future transits of the planet.

We do not consider the case with a depth of $5$\,mmag (the third
panel), an example of the first scenario, since a small secondary peak
does occur in the PPD, which therefore renders it a second scenario
case.

The figure also shows in the bottom panel the case of an undetectable
$1$\,mmag transit.  The difference in the PPD and the ITP is obvious.

\subsection{Second detection scenario}\label{1follow_up}

\begin{figure}
\includegraphics[width=0.5\textwidth]{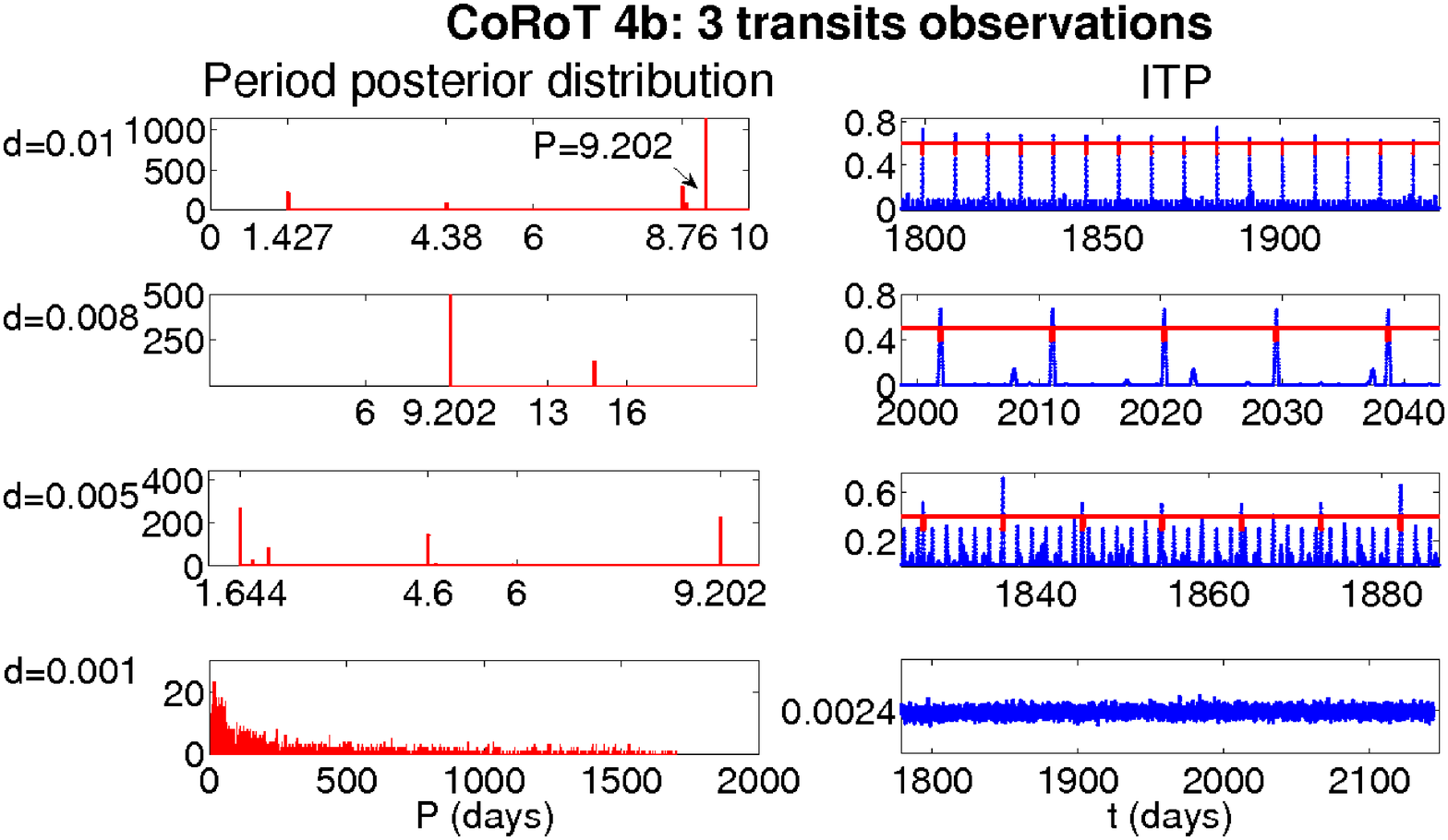}
\caption{PPDs (left panels), and ITP functions (right panels) for the
simulation of CoRoT-4b with three transit observations, for a range
of transit depths.}
\label{fig.corot4_itp}
\end{figure}

We used CoRoT-4b to demonstrate the second scenario. We chose a
transit phase that would imply sampling of three separate transits, 
although it is not very likely for such a long period planet ($P > 9\,\mathrm{d}$).
Fig.~\ref{fig.corot4_itp} shows the three cases, together with an
undetectable $1$\,mmag transit.  Clearly, the period distributions
have two or three preferred periods, that are not harmonics of each
other. That constitutes a second scenario case.  In these cases, the
most significant peaks of the ITP do coincide with the
planetary transits, so the first follow-up observation that would have
used the directed follow-up strategy, would detect it.  Again, the
bottom panel in the figure, shows, for comparison, a non-detection of
the $1$\,mmag transit.

\subsection{Third detection scenario}\label{few_follow_up}

\begin{figure}
\includegraphics[width=0.5\textwidth]{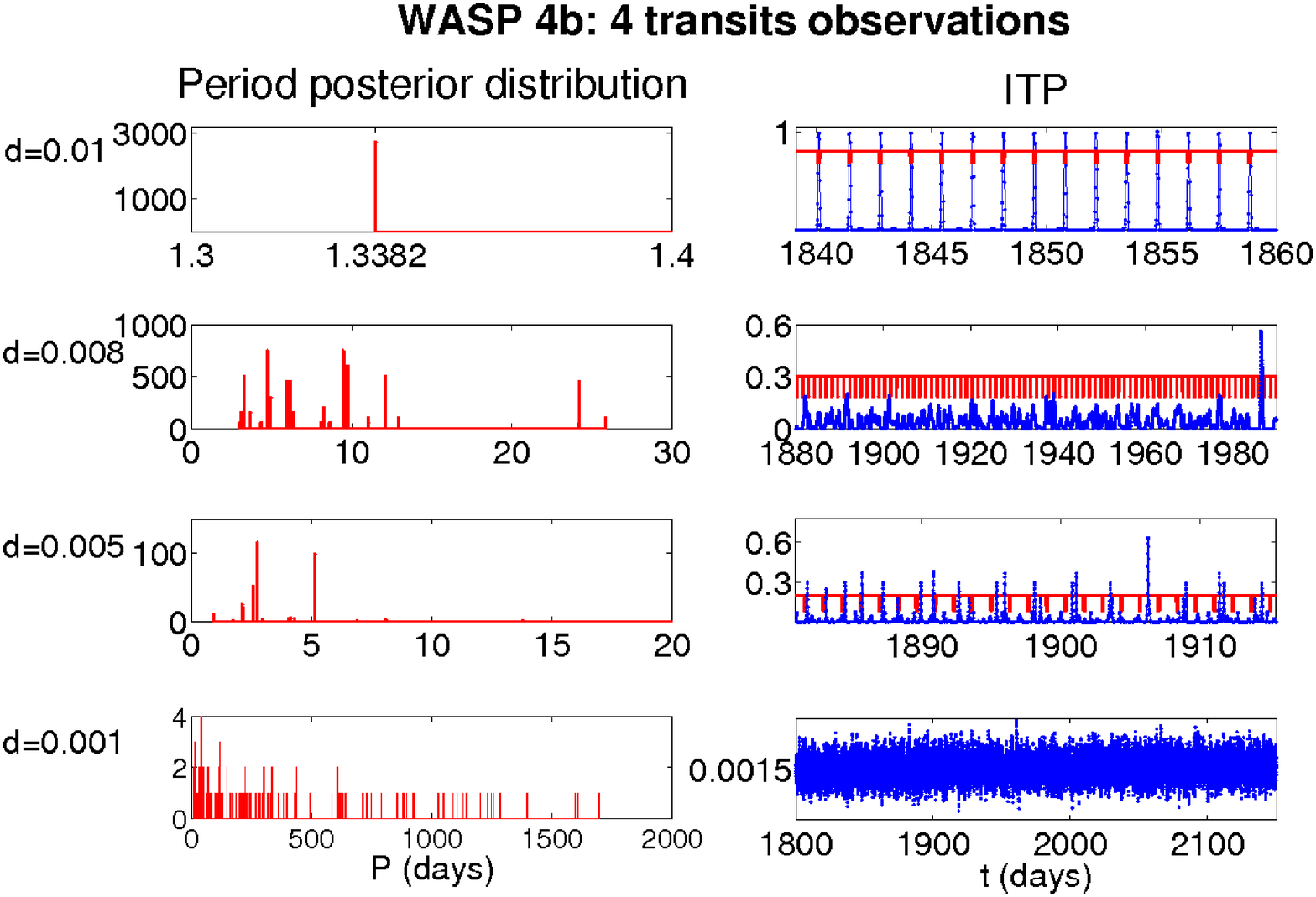}
\caption{PPDs (left panels), and ITP functions (right panels) for the
simulation of WASP-4b with four transit observations, for a range
of transit depths.}
\label{fig.wasp4_itp}
\end{figure}

The planet WASP-4b serves as the base for our third-scenario
simulations, where we chose a transit phase to sample four separate
transits.  Fig.~\ref{fig.wasp4_itp} shows the results of the
simulations with the four different transit depths we tried.  The
upper panel with a transit depth of $0.01$\,mag, is a clear example of
the first scenario, while the lower panel, with a depth of $1$\,mmag,
is a clear example of non-detection.  In the two middle panels one
sees that the period posterior distribution is multimodal.

The most prominent peaks of the resulting ITP functions do not
coincide with the actual mid-transit times. However, according to the
prioritization criteria (whose values are summarized in
Table~\ref{table.table_itp}), the star would have been prioritized for
follow-up observations. Combined with the low-cadence data, the
follow-up should eliminate some of the wrong periods, allowing others
to emerge, and eventually may lead to a detection.

\begin{figure}
\includegraphics[width=0.5\textwidth]{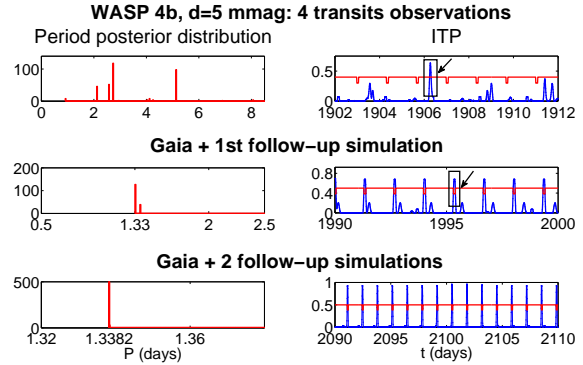}
\caption{
\textit{Top}: PPD and ITP for the \textit{Gaia} simulation of WASP-4b, with 
a depth of $5$\,mmag and four sampled transits.
\textit{Middle}: New PPD and ITP after adding the first follow-up observing sequence, that was simulated 
according to the marked ITP peak, which do not coincide with mid transit time. 
\textit{Bottom}: New PPD and ITP after adding a second follow-up sequence, that was simulated 
according to the marked ITP peak, which this time did sample a transit.}
\label{fig.wasp4_fu}
\end{figure}

We demonstrate the effect of the directed follow-up on the shallower
case, where $d=5$\,mmag. The result is depicted in
Fig.~\ref{fig.wasp4_fu}. The upper panel shows that the first
follow-up observing sequence, during the time of the highest ITP
value, happens to occur not during a transit. However, the PPD
resulting from combining the follow-up observations and \textit{Gaia}
data, modifies the distribution, as is shown in the middle panel,
together with a new suggestion for a follow-up time. A look at the
corresponding line in Table~\ref{table.table_itp} shows the
improvement in the prioritization criteria values. The next follow-up
sequence, in the bottom panel, already shows the detection.

\subsection{Mid-lifetime}\label{alert}

\begin{figure}
\includegraphics[width=0.5\textwidth]{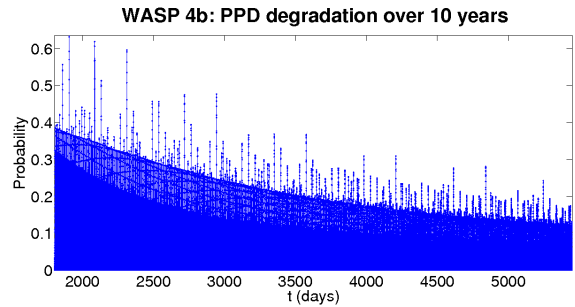}
\caption{PPD degradation rate for ten years after \textit{Gaia} will finish its operation, 
for simulation of WASP-4b, with transit depth of $5$\,mmag, and four in-transit datapoints.}
\label{fig.ppd}
\end{figure}

As we have shown in \cite{2011MNRAS.415.2513D}, as time goes by, the
ability to use the ITP to schedule follow-up observations degrades.  
We present an example of the ITP degradation rate for ten years past \textit{Gaia} in Fig.~\ref{fig.ppd}. 
It is therefore crucial that the follow-up observations will take place
as close as possible to the time the original low-cadence observations
take place. In the case of \textit{Gaia} it will probably be optimal
to perform the follow-up while the mission still operates.

\begin{figure}
\includegraphics[width=0.5\textwidth]{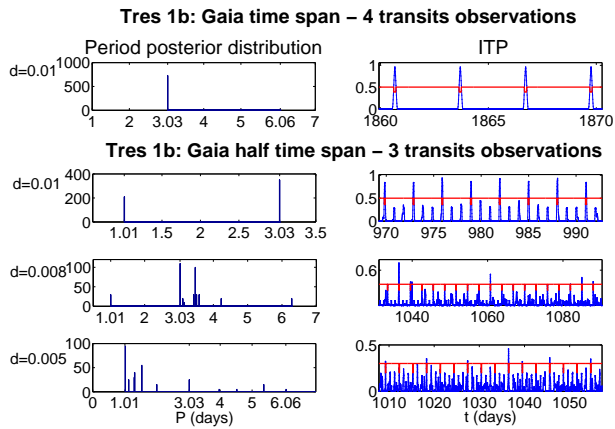}
\caption{PPD and ITP for 
the full mission lifetime simulation of TrES-1b, and for half the mission lifetime.} 
\label{fig.tres1_itp_half}
\end{figure}

We show an example of this mid-lifetime follow-up detection using the
known planet TrES-1b.  We tailored the mid-transit phase so that the
scanning law will sample four individual transits, during the full
time span of the telescope.  For a transit depth of
$0.01\,\mathrm{mag}$, the full mission lifetime yields a
first-scenario detection, as the upper panels of
Fig.~\ref{fig.tres1_itp_half} show.  The ITP peak values are
impressively close to $1$.

We then continued by simulating only half of the mission lifetime,
with the same transit phase, that implied only three observations
during transit.  According to Fig.~\ref{fig.tres1_itp_half}, the
deepest transit case, $d=0.01\,\mathrm{mag}$, results in two distinct
peaks of the PPD, which are harmonics of a single one, indicating that
this (mid-lifetime) case can be classified as a detection. The
shallower transit cases yield more than three probable periods, and
hence can be classified as third scenario cases.

The ITP values in this case are relatively high, and so is the
skewness, and we therefore conclude that there is a good chance that
the DFU strategy will prioritize this object for follow-up, even after
only half of the mission lifetime, as long as \textit{Gaia} will
sample at least three transits during this time.

\begin{figure}
\includegraphics[width=0.5\textwidth]{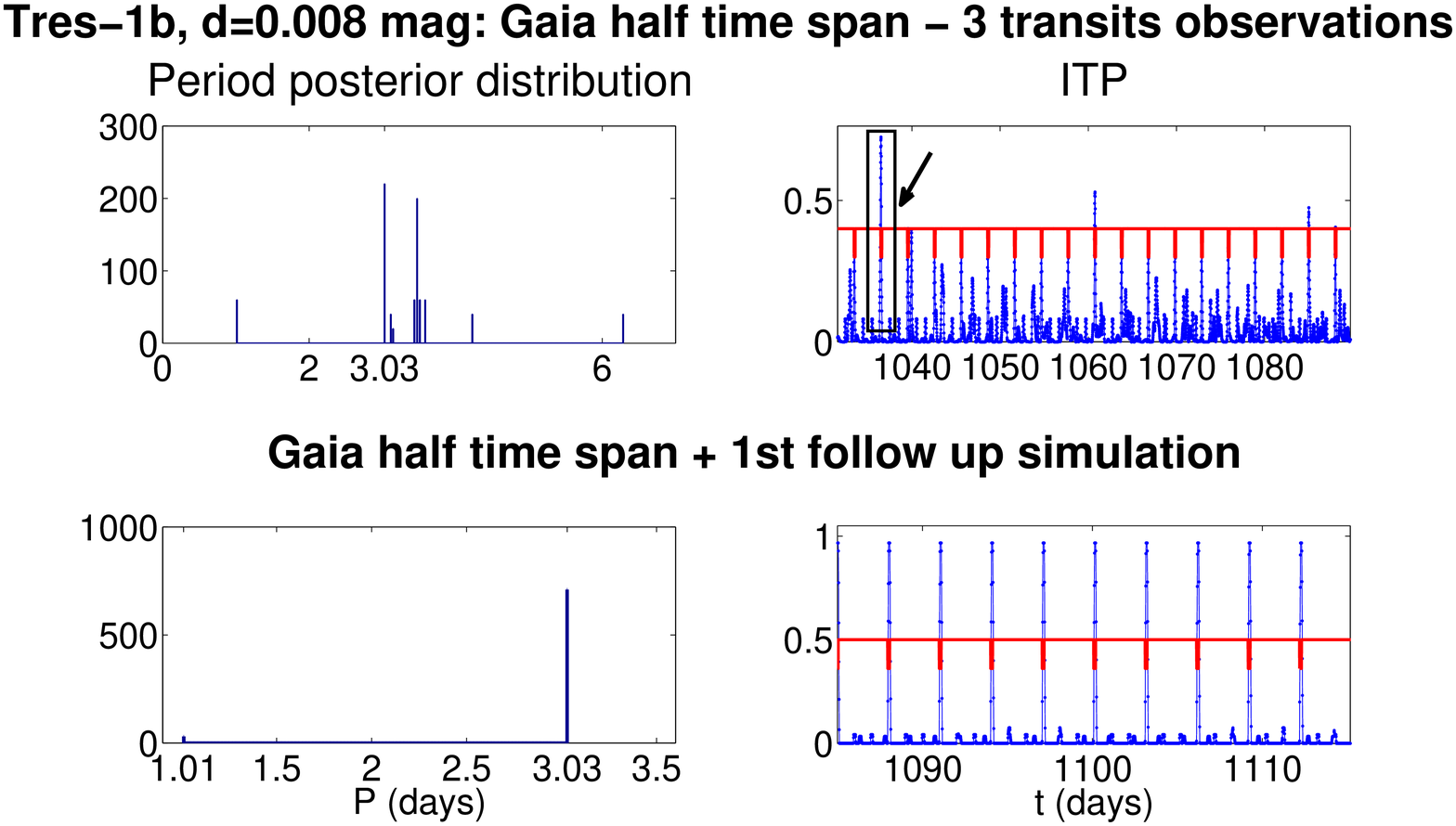}
\caption{\textit{Top}: PPD and ITP for 
the half lifetime simulation of TrES-1b with a depth of $8$\,mmag.
\textit{Bottom}: New PPD and ITP after combining the simulated follow-up observations 
(according to the marked ITP peak that coincide with mid transit) with \textit{Gaia} data.}
\label{fig.tres1_itp_half_fu2}
\end{figure}

We further simulated the follow-up observations in the case of
$d=0.008\,\mathrm{mag}$. This single observing run was enough to
exclude all the wrong periods, and allow a detection of the transiting
planet, as the bottom panels of Fig.~\ref{fig.tres1_itp_half_fu2}
show.

Based on numerous other simulations we performed of the half-lifetime
span, we conclude that we will be able to trigger follow-up
observations during \textit{Gaia}'s lifetime, and in the cases where
three transits will have been sampled in this interval, we may achieve
detections in the second or third scenario.

\subsection{False alarms}\label{false_alarm}

We chose several values for the total number of measurements
($N_{tot}=60, 70, 90, 120, \mathrm{and} 180$) and simulated $1000$
\textit{Gaia} light curves for each $N_\mathrm{tot}$, assuming white
Gaussian noise of $2-3\,\mathrm{mmag}$. We estimated the number of
simulations that would have a transit-like dimming (a double-peak
distribution of the magnitude, where the dimming would be in the range
$0.005$--$0.015\,\mathrm{mmag}$), given the increased photometric
error. We found that $28$ simulations out of $10^4$ pass this simple
test.

However, our prioritization approach excluded all those cases, since
they all had ITP skewness and peak values that were smaller than our
chosen thresholds.  The skewness values of the $28$ false-positive
cases ranged from $0.02$ to $0.76$ (three out of the $28$ had skewness
larger than $0.5$), and their highest ITP peak values were much
smaller than the $0.1$ threshold, ranging from $2\times10^{-3}$ to
$0.013$. It seems, therefore, that our prioritization scheme
effectively eliminates those false detections. 

On top of the outliers induced by white noise or stellar microvariability, calibration errors 
may also complicate the transit detection, as was examined by \cite{2011A&A..529A...6T}. 
Once \textit{Gaia} will operate, the calibration errors should be addressed 
and included in the DFU strategy, through the MCMC algorithm. 
In case ``false'' periods will be introduced to the PPD due to calibration errors, they should be eliminated by 
directed follow-up observations. 

\section{Discussion and concluding remarks}

In this paper we examined the application of the DFU strategy, which
we first introduced in \cite{2011MNRAS.415.2513D}, to \textit{Gaia}
photometry. The DFU strategy can be used to prioritize stars for
follow-up observations according to the probability to detect
transiting planets around them. We presented here selected simulated
scenarios, which represent a wider range of simulations we performed.

For all test light curves that we simulated with a transit depth
larger than $1$\,mmag, we were able to use our approach to either
recover the periodicity, or at least to propose times for directed
follow-up observations that eventually led to detection.  Furthermore,
test light curves with no transit signal were never classified as
candidates, and were not prioritized for follow-up observations. 

For transits deeper than the typical HJ depth of $0.01\,\mathrm{mag}$,
in case \textit{Gaia} will sample at least five individual transits, a
secure detection will be possible based on \textit{Gaia} data
alone. If less transits are sampled by \textit{Gaia}, the outcome of
the MH algorithm will be a multimodal PPD. In these cases we found
that we should be able to direct one or more follow-up observation
that will be scheduled according to the ITP, that will allow us to
detect the transit with minimal observational effort.

The limiting transit depth of the DFU approach, in the case of
\textit{Gaia}, seems to be around $1$\,mmag.  Simulations with this depth
were not prioritized for follow-up observations, and were
indistinguishable from pure white noise.

Since MCMC methods are computationally demanding, we first propose to
perform a simple test, in order to examine that a transit signal may
exist in the data. Only then, after choosing the stars that pass this
test, and after applying the appropriate astronomical considerations,
we propose to explore the full parameter space using the MH algorithm
and obtain the full posterior distributions. Once we run the algorithm
on the selected stars, we can continue and prioritize them for
follow-up observations according to the criteria we discussed in
section \ref{results}.


The Gaia Photometric Science Alerts Team is responsible for generating alerts on anomalous events 
that will be detected in \textit{Gaia} photometry. The facilities used by the team may also be used 
for follow-up observations, based on the DFU strategy. 
Alternatively, a dedicated follow-up network, suitable for 
observations of transiting planets (both high and low cadence) can 
complement \textit{Gaia}'s observations, and be beneficial for transit detection. 

We have noticed that the chances of detection depend directly mainly 
on the number of sampled transits, and on the transit depth. Clearly, the number of sampled transits is a 
function of the total number of measurements, as we show in Fig.~\ref{fig.window_funcion1}.  
Moreover, data points that do not sample a transit can exclude periods from the PPD, however, the 
in-transit measurements have the largest influence on the detection probability. 
This means that if, by chance, \textit{Gaia} samples
three individual transits for a certain object, before the end of the
mission, we will be able to use this partial light curve, to trigger
follow-up observations.  Obviously, the probability to sample transits
decreases as a function of decreasing number of measurements, and so
the overall yield of planets from \textit{Gaia} photometry will only
be complete once the mission ends.  Nevertheless, there is value in
starting the DFU effort while the mission still operates.  This may
expedite the detection of transiting planets based on \textit{Gaia}
photometry.

As mentioned above, we found that even for a small transit depth, of
the order of $5$\,mmag, we will usually be able to detect a planet
with five sampled transits. In case \textit{Gaia} samples less
transits, the small transit depth can yield a detection in a follow-up
campaign. The simulated case of WASP-4b with a $5$\,mmag transit,
demonstrates the prospects of detecting Neptunian planets 
around late K-stars or around M-stars.  The main
problem with small transit depths is the difficulty to perform
follow-up observations, but detections of such nature are well worth
the observational effort.

To summarize, we believe we have demonstrated in this paper the
feasibility of DFU for the case of \textit{Gaia} photometry, and
furthermore, its importance to fully exploit the extraordinary
capabilities of the mission.

\begin{table*}
\centering
  \begin{minipage}{200mm}
\caption{Known planets used in the simulations}
\label{table.planet_list}
\begin{tabular}{lccc} \hline \hline              
Planet Name & $P$ (days) & $w$ (days) & $d$ (mag) \\ \hline 
CoRoT 1b$^1$ & 1.5089557 & 0.1 & 0.025 \\
CoRoT 4b$^2$ & 9.20205 & 0.1583 & 0.013 \\
TrES 1b$^3$ & 3.0300722 & 0.104 & 0.022 \\
WASP 4b$^4$ & 1.3382282 & 0.0928 & 0.02 \\
\hline \hline             
    \end{tabular}
\newline
References:\\
 1. \cite{2008A&A...482L..17B}
 2. \cite{2008A&A...488L..43A}
 3. \cite{2009AN....330..475R}
 4. \cite{2008ApJ...675L.113W}
\end{minipage}
\end{table*}

\begin{table*}
\centering
  \begin{minipage}{200mm}
\caption{Summary of the simulation results}
\label{table.table_itp}

   \begin{tabular}{lccccccc} \hline \hline              
planet name& \ $N_\mathrm{tot}$  & $N_\mathrm{tr}$ & $d$  &  $\langle d \rangle$ & $W$  & Maximum ITP &  $S$ \\ \hline  
CoRoT 1b &64& 5 & 0.01& 0.0099 & 70.9 & 1 & 2.5  \\
CoRoT 1b &64& 5 & 0.008 &0.0081& 205.2 & 1 & 4.88   \\
CoRoT 1b &64& 5 & 0.005 & 0.0051& 44 & 1 & 4.16  \\
CoRoT 1b &64& 5 & 0.001 & 0.075 & 3.4 & $2.1\times10^{-3}$ & 0.02  \\
CoRoT 4b &63& 3  & 0.01 &0.0094 & 80.33 & 0.74 & 5.93   \\
CoRoT 4b &63& 3 & 0.008&0.0076 & 20.1 & 0.75 & 5.39   \\
CoRoT 4b &63& 3 & 0.005&0.0048 & 10.46 & 0.71 & 2.5   \\
CoRoT 4b &63& 3 & 0.001    &0.08   & 2.2 & $2.5\times10^{-3}$ & 0.08   \\
WASP 4b &83& 4 & 0.01 &0.01&82.2&0.98 &3.34 \\
WASP 4b &83& 4 & 0.008 &0.0079&90.1&0.57&2.17 \\
WASP 4b &83& 4 & 0.005 &0.005&79.2&0.63&3.5 \\
WASP 4b &83& 4 & 0.001 &0.02&7.1&$1.9\times10^{-3}$&0.08 \\
WASP 4b + 1st FU &83+4& 4 & 0.005 &0.005&180.4&0.7&2.86 \\
WASP 4b + two FU &83+4+4& 4+1 & 0.005 &0.005&181.5&0.96&3.32 \\
TrES 1b &98& 4 & 0.01 &0.01&19.6&0.97 &4.5 \\
TrES 1b &48& 3 & 0.01 &0.0098&83.4&0.94&3.1 \\
TrES 1b &48& 3 & 0.008 &0.008&47&0.72&2.67 \\
TrES 1b &48& 3 & 0.005 &0.0048&21.8&0.46&1.6 \\
TrES 1b + 1st FU &48+4& 3+1 & 0.008 &0.0078&41.8&0.97&4.4 \\
\hline \hline             

    \end{tabular}
\end{minipage}
\end{table*}

\label{lastpage}

\end{document}